\documentclass[aps,twocolumn,preprintnumbers,groupedaddress,superscriptaddress,floatfix,
tightenlines,reprint,nofootinbib,longbibliography]{revtex4-1}
\usepackage{mathrsfs}
\usepackage{natbib}
\usepackage{verbatim}
\usepackage{enumerate}
\usepackage{graphicx,bbm,amsbsy,amsfonts,amssymb,amsmath}
\usepackage{bm}
\usepackage{hyperref}
\usepackage{booktabs}
\usepackage{gensymb}

\usepackage{footmisc}

\usepackage{blindtext}

\def\2{{(2)}}
\def\1{{(1)}}
\def\0{{(0)}}
\def\m1{{(-1)}}

\let\oldAA\AA
\renewcommand{\AA}{\text{\normalfont\oldAA}}

\hypersetup{
    colorlinks=true,       % false: boxed links; true: colored links
    linkcolor=red,          % color of internal links (change box color with linkbordercolor)
    citecolor=blue,        % color of links to bibliography
    filecolor=magenta,      % color of file links
    urlcolor=blue           % color of external links
}

\newcommand{\be}{\begin{equation}}
\newcommand{\ee}{\end{equation}}
\newcommand{\ea}{\end{eqnarray}}
\newcommand{\ba}{\begin{eqnarray}}
\begin{document}

%%%%%%%%%%%%%%%%%%%%%%%%%%%%%%%%%%%%%%%%%%%%%%%%%%%%%%%%%%%%%%%%%%%%%%%%%%%%%%%%%%%%%%%%%%%%%%%%%%%%

%\preprint{UMN–TH–XXXX/XX}
%\preprint{FTPI–MINN–XX/XX}

\title{Non-invertible Chiral Symmetry and Axions under Electromagnetic Duality}

\author{Gongjun Choi}
\email{choi0988@umn.edu}
\affiliation{William I. Fine Theoretical Physics Institute, School of Physics and Astronomy,\\
University of Minnesota, Minneapolis, Minnesota 55455, USA}

\author{Tony Gherghetta}
\email{tgher@umn.edu}
\affiliation{School of Physics and Astronomy, University of Minnesota, Minneapolis, Minnesota 55455, USA}

\author{John Terning}
\email{jterning@gmail.com}
\affiliation{Center for Quantum Mathematics and Physics (QMAP),\\Department of Physics, University of California, Davis, California 95616, USA}

%%%%%%%%%%%%%%%%%%%%%%%%%%%%%%%%%%%%%%%%%%%%%%%%%%%%%%%%%%%%%%%%%%%%%%%%%%%%%%%%%%%%%%%%%%%%%%%%%%%%

\begin{abstract}
We study the implications of non-invertible chiral symmetry in a four-dimensional U(1) gauge theory coupled to massless fermions with electromagnetic $SL(2,\mathbb{Z})$ duality. This is done by deriving the Adler-Bell-Jackiw anomaly of massless QED in the dual frame that is used to explicitly construct the symmetry defect operator as well as the conserved two-form symmetry current. As expected, the non-invertible chiral symmetry is covariant under the duality transformation. This has implications for understanding the nature of kinetic and topological terms in the dual frame and for axion electrodynamics. In particular, we show that to generate an axion potential from a dyon loop, the one-form magnetic symmetry must be explicitly broken by a mutually non-local charged state with nonzero pairwise helicity.

\end{abstract}

%%%%%%%%%%%%%%%%%%%%%%%%%%%%%%%%%%%%%%%%%%%%%%%%%%%%%%%%%%%%%%%%%%%%%%%%%%%%%%%%%%%%%%%%%%%%%%%%%%%%

\date{\today}
\maketitle

%%%%%%%%%%%%%%%%%%%%%%%%%%%%%%%%%%%%%%%%%%%%%%%%%%%%%%%%%%%%%%%%%%%%%%%%%%%%%%%%%%%%%%%%%%%%%%%%%%%%

\section{Introduction}

Symmetries play a central role in quantum field theory, governing both the dynamics and classification of possible phases. Recently, there has been significant progress in generalizing our notions of symmetry that deepens our understanding of quantum field theory. These new global symmetries are reformulated by introducing topological and gauge invariant symmetry defect operators~\cite{Gaiotto:2014kfa}. The conservation law of these generalized global symmetries is then captured by the invariance of the symmetry defect operator under the deformation of a supporting manifold, revealing the topological nature of the symmetry defect operator.

Alongside higher-form global symmetries~\cite{Gaiotto:2014kfa,Aharony:2013hda,Gomes:2023ahz,Bhardwaj:2023kri}, non-invertible symmetries form an integral part of generalized global symmetries (see ~\cite{Schafer-Nameki:2023jdn,Shao:2023gho,Brennan:2023mmt} and references therein). The non-invertible symmetry operation, unlike the usual invertible symmetry, obeys a more general fusion rule where the Hermitian conjugate is not an inverse symmetry operation on external extended objects that can be used to probe the theory. One of the simplest examples of a non-invertible symmetry is associated with the chiral symmetry 
in a 4D Abelian gauge theory that has an Adler-Bell-Jackiw (ABJ) anomaly~\cite{Adler:1969gk,Bell:1969ts}. In particular, for massless quantum electrodynamics (QED), a non-invertible chiral symmetry defect can be constructed for a chiral rotation~\cite{Choi:2022jqy,Cordova:2022ieu} (see also \cite{Damia:2022bcd,Karasik:2022kkq,GarciaEtxebarria:2022jky,Yokokura:2022alv,Choi:2022fgx,vanBeest:2023dbu,Benedetti:2023owa,Arbalestrier:2024oqg,Chen:2025buv,Gagliano:2025oqv}). This is made possible by coupling the theory to a $Z_N$ topological quantum field theory (TQFT) on codimension-one defects, which gauges a discrete subgroup of the magnetic one-form symmetry in the bulk. The intriguing non-perturbative physics underlying the breaking of the non-invertible symmetry can be useful in explaining hierarchies that  often appear in particle physics, and for this reason, non-invertible chiral symmetry has found increasing applications in physics beyond the Standard Model~\cite{Cordova:2022ieu,Cordova:2022fhg,Cordova:2023her,Cordova:2024ypu,Delgado:2024pcv}.

Aside from the selection rules imposed by the non-invertible chiral symmetry, it is also interesting to consider how the construction of symmetry defect operators behaves under dualities. Since duality transformations, such as electromagnetic duality in Abelian gauge theories, are expected to preserve selection rules imposed by symmetries, any non-invertible chiral symmetry present in one frame must have a counterpart in the dual frame. In the case of massless QED, this motivates the search for a non-invertible chiral symmetry in the $SL(2,\mathbb{Z})$ dual frame—specifically, in a $U(1)$ gauge theory coupled to massless fermionic dyons.

Electromagnetic duality in Abelian gauge theories \cite{Cardy:1981qy,Cardy:1981fd,Shapere:1988zv}, parameterized by the modular group $SL(2,\mathbb{Z})$, exchanges electric and magnetic fields while mixing the $\theta$ angle and the gauge coupling~\cite{Deser:1976iy,Witten:1995gf}. Under such a transformation, electrically-charged fermions are mapped to dyons that have both electric and magnetic charge. While the duality is manifest in the classical Maxwell equations, implementing it at the quantum level \cite{Colwell:2015wna} requires careful tracking of anomalies and defect operators. In the dual frame, the standard $U(1)$ current algebra is modified due to the dyonic nature of the fermions, and the ABJ anomaly takes a more intricate form, mixing both $F \wedge F$ and $F \wedge \star F$ structures.

In this paper, we explicitly construct the non-invertible chiral symmetry defect in a $U(1)$ gauge theory with massless fermionic dyons, confirming its presence in the $SL(2,\mathbb{Z})$ dual of massless QED. Starting from the known non-invertible symmetry in the QED frame, and using the $SL(2,\mathbb{Z})$ transformation, we track the effect of a chiral rotation on the complexified coupling $\tau$ and derive the ABJ anomaly in the dual frame \cite{Csaki:2010rv}. We show that this anomaly can be written 
as the wedge product of the Hodge dual of a conserved current of the  one-form symmetry, which is therefore a two-form, enabling the construction of a topological defect operator that implements the non-invertible chiral symmetry. The current of the one-form symmetry in the bulk couples to a $Z_N$ one-form gauge field via a Chern–Simons interaction, consistent with the known fractional quantum Hall interpretation of such defects.

Our construction confirms that the existence of non-invertible chiral symmetry is not contingent on the choice of a duality frame, and it provides a concrete realization of how the non-invertible chiral symmetry can persist in theories with dyonic matter. Throughout the paper, we will designate  the {\it QED frame} to be the frame with the standard Abelian ABJ anomaly 
\be
{\rm d}\star j_{\rm chiral}=\frac{\mathcal{A}}{8\pi^{2}}F\wedge F\,, 
\ee
where $j_{\rm chiral}$ is the anomalous $U(1)_{\rm chiral}$ current of electrically charged fermions,
$F$ is the holomorphically normalized gauge field strength and $\mathcal{A}\in\mathbb{Z}$ is an anomaly coefficient. An $SL(2,\mathbb{Z})$ duality transformation performed on the QED frame will give rise to a new frame, which will be referred to as the {\it dual frame}.

After we review the properties of the Abelian gauge theory in the QED frame, and how those properties transform under $SL(2,\mathbb{Z})$ in Sec.~\ref{sec:electronframe} and \ref{sec:SL2Z}, respectively, we derive the ABJ anomaly in the dual frame in Sec.~\ref{sec:ABJanomaly}. In Sec.~\ref{sec:INsymdyon}, we demonstrate the presence of the non-invertible chiral symmetry in the dual frame by specifying the counterpart of the current generating the one-form magnetic symmetry in the QED frame and its coupling to the minimal $Z_{N}$ TQFT. In Sec.~\ref{sec:implications}, we further explore implications of this construction. We study how the non-invertible chiral rotation acts on the theory at a fixed point in the renormalization group flow, where the electric and magnetic contributions to the photon self-energy cancel. We demonstrate that the $U(1)_{\rm chiral}$ rotation does not spoil the fixed point in the dual frame, in spite of the exotic anomalous Ward identity ${\rm d}\star j_{\rm chiral}\propto F\wedge\star F$ in the dual frame. Furthermore, we revisit axion electrodynamics in the presence of massless dyons. We discuss the selection rule on an axion potential imposed by the non-invertible chiral  symmetry in the dual frame of axion-Maxwell theory. Finally, in Appendix~\ref{sec:AppendixA}, we derive the renormalization group equation for the complexified gauge coupling in the dual frame and in Appendix~\ref{sec:AppendixB} we derive the transformation of a dyon line operator under the non-invertible chiral symmetry.

%%%%%%%%%%%%%%%%%%%%%%%%%%%%%%%%%%%%%%%%%%%%%%%%%%%%%%%%%%%%%%%%%%

\section{Overview of QED and EM duality}

In this section we briefly review massless QED, in order to define and introduce our notation.

\subsection{Aspects of QED}\label{sec:electronframe}

Consider a $U(1)$ gauge theory with $N_{f}$ massless Dirac fermions $\Psi_{j}$ ($j=1,\dots, N_{f}$) in the QED frame. The action of the theory in Minkowski space reads
\ba
S&=&\int-\frac{1}{2e^{2}}F\wedge\star F+\frac{\theta}{8\pi^{2}}F\wedge F-A\wedge\star J\cr\cr
&&+\int d^{4}x\sqrt{-g}\,\,\sum_{j=1}^{N_{f}}i\bar{\Psi}_{j}\partial_{\mu}\gamma^{\mu}\Psi_{j}\,,
\label{eq:Selectron}
\ea
where $e$ is the gauge coupling, $\theta$ is a $2\pi$-periodic angle, and $g$ is the metric determinant. The gauge sector of the action is written in differential form for later convenience where $F\equiv{\rm d}A$ is the globally defined gauge field strength, $\star F$ is the Hodge dual of $F$, and $\star J$ is the electric current. For discussions involving magnetic charges or currents, we shall denote the corresponding magnetic current as $\star K$, keeping in mind that the Lagrangian (\ref{eq:Selectron}) describes only the electrically charged fields.\footnote{A Lorentz-invariant, local Lagrangian description of Maxwell theory coupled to magnetic matter can be obtained by gauging the electric one-form symmetry $U(1)^{[1]}_{e}$ with the aid of auxiliary gauge fields associated with $U(1)^{[1]}_{e}$ and its dual 0-form $U(1)$ symmetry; see Sec.~5.3 in \cite{Chen:2025buv}.}

Away from electric charge or current $({\rm i.e.}~\star J=0)$, and where the $\theta$ term is not a total derivative (e.g. due to the presence of magnetically charged objects), the equation of motion of $A$ arising from \eqref{eq:Selectron} is
\be
{\rm d}\left(\frac{1}{e^{2}}\star F-\frac{\theta}{2\pi}\frac{F}{2\pi}\right)=0\,.
\label{eq:Aeom}
\ee
Because any closed form is exact locally, we can introduce the dual gauge field $\tilde{A}$ to rewrite the expression in the parenthesis of \eqref{eq:Aeom} as\footnote{Note that when the gauge field is canonically normalized, the minimal coupling becomes $-eA\wedge\star J-\tilde{A}\wedge\star K$ with 
\be
\frac{{\rm d}\tilde{A}}{2\pi}=-\star F+e^{2}\frac{\theta}{2\pi}\frac{F}{2\pi}\,.\nonumber
%\label{eq:dualA2}
\ee 
} 
\be
\frac{{\rm d}\tilde{A}}{2\pi}=-\frac{1}{e^{2}}\star F+\frac{\theta}{2\pi}\frac{F}{2\pi}\,.
\label{eq:dualA}
\ee 
If the exterior derivative of the second term on the right hand side of (\ref{eq:dualA}) vanishes (e.g. away from a magnetically charged object), it simply drops out in the dual gauge field definition. Although $\tilde{A}$ is not globally defined, we can still patch together dual gauge fields defined locally in different open patches of spacetime as long as the $\tilde{A}$ in the different coordinate patches are gauge-equivalent on the patch overlaps.

In regions where $\star J\neq0$ and the $\theta$ term is not a total derivative, the equation of motion of $A$ is now given by
\be
{\rm d}\left(\frac{1}{e^{2}}\star F-\frac{\theta}{2\pi}\frac{F}{2\pi}\right)=\star J\,.
\label{eq:currentelectric}
\ee
When Eq.~(\ref{eq:currentelectric}) is integrated over a 3-manifold $\Sigma$ in spacetime with a boundary $\partial \Sigma\neq\varnothing$, we obtain the usual integrally-quantized electric charge
\be
Q_{\rm e}=\int_{\Sigma}\star J=\int_{\partial \Sigma}\left(\frac{1}{e^{2}}\star F-\frac{\theta}{2\pi}\frac{F}{2\pi}\right)\in\mathbb{Z}\,,
\label{eq:Qe}
\ee 
where Stoke's theorem has been used in the second equality of \eqref{eq:Qe}. When the magnetic flux is activated by the presence of the magnetic current $\star K$, we have
\be
{\rm d}\left(\frac{F}{2\pi}\right)=\star K\,.
\label{eq:unclosedF}
\ee
Similarly, when Eq.~(\ref{eq:unclosedF}) is integrated over a 3-manifold $\Sigma$ with a boundary $\partial \Sigma\neq\varnothing$, we obtain integrally-quantized magnetic charge $Q_{\rm m}$
\be
Q_{\rm m}=\int_{\Sigma}\star K=\int_{\partial \Sigma}\frac{F}{2\pi}\in\mathbb{Z}\,.
\label{eq:Qm}
\ee
As the magnitude of the magnetic charge is integrally quantized, we take $2\pi/e$ as the unit of magnetic charge throughout this work. When $\theta\neq0$, since the $\theta$-dependent contribution in (\ref{eq:Qe}) evaluates to the magnetic charge in (\ref{eq:Qm}), a magnetic monopole with magnetic charge $Q_{\rm m}$ obtains a $\theta$-dependent electric charge, which is nothing but the Witten effect~\cite{Witten:1979ey}. In this way, a magnetically charged object becomes a dyon.

Note that if there are no dynamical charged fields in the QED frame, there are {\it two} one-form global symmetries that act on line operators. These are the electric one-form  symmetry,  $U(1)^{[1]}_{e}$, and the magnetic one-form symmetry, $U(1)^{[1]}_{m}$,  which have corresponding two-form currents: $j_{e}^{[2]}=\frac{F}{e^{2}}$ and $j^{[2]}_{m}=\frac{\star F}{2\pi}$, respectively~\cite{Gaiotto:2014kfa}. The electric one-form symmetry acts on the Wilson line $W_{q_{e}}(\gamma_{e})\equiv{\rm exp}(iq_{e}\int_{\gamma_{e}}A)$, where $\gamma_{e}$ is the path of a probe charge $q_e$, by shifting $A$ by a flat connection. On the other hand, the operator charged under the magnetic one-form symmetry $U(1)_{m}^{[1]}$ is a `t Hooft line, denoted by $T_{q_{m}}(\gamma_{m})$, which corresponds to the worldline $\gamma_{m}$ of an infinitely heavy magnetic monopole of charge $q_{m}$. 

When we move $\gamma_{e}$ around $\gamma_{m}$ or vice versa, there arises a non-trivial mutual phase 
between the line operators that results in a nontrivial commutation relation~\cite{Aharony:2013hda} 
\be
T_{q_{m}}(\gamma_{m})W_{q_{e}}(\gamma_{e})=e^{2\pi iq_{e}q_{m}}W_{q_{e}}(\gamma_{e})T_{q_{m}}(\gamma_{m})\,.
\label{eq:linecommute}
\ee
The phase in \eqref{eq:linecommute} is similar to the Aharonov-Bohm effect and reflects the mutual non-locality between the electric and magnetic charges. However, to ensure consistency of the quantum theory, locality must be preserved and correlation functions involving Wilson and ’t Hooft line operators must be single-valued. Hence, this requires the phase factor in \eqref{eq:linecommute} to be ${\rm exp}[2\pi iq_{e}q_{m}]=1$, which yields the Dirac quantization condition $q_{e}q_{m}\in\mathbb{Z}$ (in our charge convention).

%Although still a local QFT, the theory cannot be described by a manifestly Lorentz-invariant, local Lagrangian with a single gauge field that couples to both electrically and magnetically-charged fermions. 
Although the theory cannot be described by a manifestly Lorentz-invariant, local Lagrangian with a single gauge field that couples to both electrically and magnetically-charged fermions, it is still a local QFT with charged particle states.
%However, since this is still a local QFT we can consider two fields, labeled 1 and 2, with both electric and magnetic charges.
%This allows us to study the particle states. 
In particular, consider two fields, labeled 1 and 2, with both electric and magnetic charges.
%Consider a theory with two charged fields, labeled 1 and 2. Then for a state with one of each type of particle, 
Then, if the pairwise helicity \cite{Csaki:2020inw,Csaki:2020yei} (or Dirac-Schwinger-Zwanziger antisymmetric product), defined as
\be
\tfrac{1}{2}(Q_{\rm e1}Q_{\rm m2}-Q_{\rm e2}Q_{\rm m1})~,
\label{eq:pairwise}
\ee 
is non-vanishing, the fields are said to be {\it mutually non-local}. Note that (\ref{eq:pairwise}) is equivalent to the magnitude of the angular momentum stored in the electromagnetic fields and thus is half-integrally quantized due to the quantization of angular momentum. When there are more than two matter fields, the pairwise helicity can be defined for each pair of fields. Crucially, the pairwise helicity \eqref{eq:pairwise} is invariant under $SL(2,\mathbb{Z})$ as will be discussed in Sec.~\ref{sec:SL2Z}.

%%%%%%%%%%%%%%%%%%%%%%%%%%%%%%%%%%%%%%%%%%%%%%%%%%%%%%%%%%%%%%%%%%

\subsection{$SL(2,\mathbb{Z})$ duality}
\label{sec:SL2Z}

We next review electromagnetic $SL(2,\mathbb{Z})$ duality. This duality transformation will be used in subsequent sections to study the ABJ anomaly and non-invertible chiral symmetry. Consider the complexified gauge coupling\footnote{Due to the factor $2\pi$ in the imaginary part, the minimum magnitude of a non-zero magnetic charge is 1.}
\be
\tau\equiv\frac{\theta}{2\pi}+\frac{2\pi}{e^{2}}i\,.
\label{eq:tau}
\ee
The particle spectrum of the theory is invariant under $\tau\rightarrow\tau+\mathbb{Z}$ (T-duality) and $\tau\rightarrow-\tau^{-1}$ (S-duality). Iterations of these transformations gives the full $SL(2,\mathbb{Z})$ group under which $\tau$ transforms as
\be
\tau\rightarrow\frac{a\tau+b}{c\tau+d}\,,
\label{eq:tauSL}
\ee
where $a,b,c,d$ are constants satisfying $ad-bc=1$. Similarly, the respective transformations of $(A,\tilde{A})$ and $(J,K)$ are
\be
\begin{bmatrix}
    d       & c\\
    b       & a
\end{bmatrix}\begin{bmatrix}
    A      \\
    \tilde{A}    %  
\end{bmatrix}=\begin{bmatrix}
    A'      \\
    \tilde{A}'   %  
\end{bmatrix}\,,
\label{eq:ASL}
\ee
and
\be
\begin{bmatrix}
    a       & -b\\
    -c       & d
\end{bmatrix}\begin{bmatrix}
    \star J      \\
    \star K      
\end{bmatrix}=\begin{bmatrix}
    \star J'      \\
    \star K'     
\end{bmatrix}\,,
\label{eq:JSL}
\ee
where prime ($^\prime$) is used to denote the original QED frame while unprimed notation is reserved for the dual frame. Also, (\ref{eq:JSL}) implies 
\be
\begin{bmatrix}
    a       & -b\\
    -c       & d
\end{bmatrix}\begin{bmatrix}
    Q_{\rm e}      \\
    Q_{\rm m}      
\end{bmatrix}=\begin{bmatrix}
    Q'_{\rm e}      \\
    Q'_{\rm m}     
\end{bmatrix}\,.
\label{eq:QSL}
\ee
One can easily see that (\ref{eq:QSL}) is consistent with the requirement that the electric charge of a dyon just shifts by an integer under a T-duality transformation. 
Particularly when 
\be
c=Q_{\rm m}/n\,, \quad\quad d=Q_{\rm e}/n\,,
\label{dualityparam}
\ee
with $n={\rm gcd}(Q_{\rm e},Q_{\rm m})$, the electric and magnetic charges in the dual frame, $Q_{\rm e}$ and $Q_{\rm m}$, are mapped to $Q'_{\rm e}=n$ and $Q'_{\rm m}=0$ in the QED frame. Furthermore, using (\ref{eq:QSL}), we can explicitly check the invariance of the pairwise helicity (\ref{eq:pairwise}) under $SL(2,\mathbb{Z})$, which follows from
\ba
&&\begin{bmatrix}
    Q'_{\rm e2},\,
    Q'_{\rm m2}      
\end{bmatrix}\begin{bmatrix}
    0       & -1\\
    1       & 0
\end{bmatrix}\begin{bmatrix}
    Q'_{\rm e1}      \\
    Q'_{\rm m1}      
\end{bmatrix}\cr\cr
&&=\begin{bmatrix}
    Q_{\rm e2},\,
    Q_{\rm m2}      
\end{bmatrix}\begin{bmatrix}
    a       & -b\\
    -c       & d
\end{bmatrix}^{\rm T}\begin{bmatrix}
    0       & -1\\
    1       & 0
\end{bmatrix}\begin{bmatrix}
    a       & -b\\
    -c       & d
\end{bmatrix}\begin{bmatrix}
    Q_{\rm e1}      \\
    Q_{\rm m1}      
\end{bmatrix}\,,\cr\cr
&&=\begin{bmatrix}
    Q_{\rm e2},\,
    Q_{\rm m2}      
\end{bmatrix}\begin{bmatrix}
    0       & -1\\
    1       & 0
\end{bmatrix}\begin{bmatrix}
    Q_{\rm e1}      \\
    Q_{\rm m1}      
\end{bmatrix}\,.
\label{eq:pairwisehelicityinv}
\ea

Given (\ref{eq:tau}), we can write (\ref{eq:currentelectric}) and (\ref{eq:unclosedF}) in a more compact form
\be
\frac{{\rm Im}[\tau]}{2\pi}({\rm d}\star F+i{\rm d}F)=\star J+\tau\star K\,.
\label{eq:combinedMW}
\ee
Because (\ref{eq:tauSL}) and (\ref{eq:JSL}) are equivalent to the transformation rule
\be
\frac{\star J+\tau\star K}{c\tau+d}=\star J'+\tau' \star K'\,,
\label{eq:eqmSL}
\ee
the covariance of (\ref{eq:combinedMW}) under $SL(2,\mathbb{Z})$ implies~\cite{Csaki:2010rv}
\be
(c\tau^{*}+d)({\rm d}\star F+i{\rm d}F)={\rm d}\star F'+i{\rm d}F'\,.
\label{eq:Ftransf}
\ee

Given the transformation rules for $(A,\tilde{A})$ in (\ref{eq:ASL}) and for $(F,\star F)$ in (\ref{eq:Ftransf}), one may wonder if these two rules are compatible. To check this, we first identify the imaginary part of both sides in (\ref{eq:Ftransf}), which gives
\ba
&&F'=-c\left(\frac{2\pi}{e^{2}}\right)\star F+\left(c\frac{\theta}{2\pi}+d\right)F\cr\cr
\Leftrightarrow &&A'=dA+c\tilde{A}\,,
\label{eq:Aprime}
\ea
where the definition (\ref{eq:dualA}) of the dual gauge field $\tilde{A}$ has been used to obtain the second line in \eqref{eq:Aprime}. On the other hand, using (\ref{eq:dualA}) and the field strength transformation rules in (\ref{eq:tauSL}) and (\ref{eq:Ftransf}), we find 
\ba
{\rm d}\tilde{A}'&=&{\rm Im}[\tau^{*'}(\star F'+iF')]\,,\cr\cr
&=&{\rm Im}[(a\tau^{*}+b)(\star F+iF)]\,,\cr\cr
&=&a\left(-\frac{2\pi}{e^{2}}\star F+\frac{\theta}{2\pi}F\right)+bF\,,
\ea
which shows that $\tilde{A}'=bA+a\tilde{A}$, in agreement with (\ref{eq:ASL}).

%%%%%%%%%%%%%%%%%%%%%%%%%%%%%%%%%%%%%%%%%%%%%%%%%%%%%%%%%%%%%%%%%%

\section{Aspects of Chiral symmetry in the dual frame}

In this section, we consider how the anomalous $U(1)_{\rm chiral}$ symmetry is manifested in the dual frame. We first discuss the ABJ anomaly in the dual frame and then show the presence of the non-invertible chiral symmetry in the dual frame by identifying the conserved two-form current.

\subsection{ABJ anomaly in the dual frame}
\label{sec:ABJanomaly}

To reveal how the ABJ anomaly appears in the dual frame we can compare the complex gauge couplings $\tau$ and $\tau_{\{\alpha\}}$ in the dual frame before and after we perform the chiral rotation, respectively. These couplings are obtained by an $SL(2,\mathbb{Z})$ transformation of the corresponding couplings in the QED frame.

Consider first the gauge action in the dual frame
\ba
S&\supset&\int-\frac{1}{2e^{2}}F\wedge\star F+\frac{\theta}{8\pi^{2}}F\wedge F+...\cr\cr
&=&\int-\frac{{\rm Im}[\tau]}{4\pi}F\wedge\star F+\frac{{\rm Re}[\tau]}{4\pi}F\wedge F+...\,,
\label{eq:Sbefore}
\ea
where the complexified gauge coupling in the dual frame before the chiral rotation is given by \eqref{eq:tau}. In the QED frame, this coupling corresponds to $\tau'$ which can be obtained by applying (\ref{eq:tauSL}) to give
\be
\tau'=\frac{a\tau+b}{c\tau+d}\,.
\label{eq:tauprimebefore}
\ee
When we perform the chiral rotation of $N_{f}$ Dirac fermions, $\Psi_{j}=(\psi_{j},\xi_{j}^{*})^{T}$ with $j=1\dots N_{f}$ by an angle $\alpha\in[0,2\pi]$, i.e. $\psi_{j}\rightarrow e^{i\alpha}\psi_{j}$ and $\xi_{j}\rightarrow\xi_{j}$, there appears a shift of $\theta'=2\pi{\rm Re}[\tau']$ by an amount
\be
\frac{{\rm Re}[\tau']}{4\pi}F'\wedge F'
\rightarrow\underbrace{\left({\rm Re}[\tau']-\frac{\mathcal{A}\alpha}
{2\pi}\right)}_{{\rm Re}[\tau_{\{\alpha\}}']}\frac{1}{4\pi}F'\wedge F'\,,
\label{eq:Schangeelectronframe}
\ee
where $\mathcal{A}=\sum_{j=1}^{N_{f}}Q_{{\rm chiral},j}Q_{{\rm e},j}^{2}\in\mathbb{Z}$ is the anomaly coefficient.

Since only the real part changes in the QED frame, we can write the complexified gauge coupling after the $U(1)_{\rm chiral}$ rotation as
\ba
\tau_{\{\alpha\}}'\equiv \tau'-\frac{\mathcal{A}\alpha}{2\pi}
= \frac{a\tau+b}{c\tau+d}-\frac{\mathcal{A}\alpha}{2\pi}\,.
\label{eq:tauNf}
\ea
where (\ref{eq:tauprimebefore}) was used in the second equality.

Finally to obtain $\tau_{\{\alpha\}}$ in the dual frame, we apply the inverse of (\ref{eq:tauSL}) to (\ref{eq:tauNf}) to obtain
\ba
\tau_{\{\alpha\}}&=&\frac{d\tau_{\{\alpha\}}'-b}{-c\tau_{\{\alpha\}}'+a}=\frac{\tau-\mathcal{A}d(c\tau+d)\frac{\alpha}{2\pi}}{1+\mathcal{A}c(c\tau+d)\frac{\alpha}{2\pi}}\,,\nonumber\\
&=&\tau-\mathcal{A}(c\tau+d)^2\frac{\alpha}{2\pi}+\mathcal{O}\left[\left(\frac{\alpha}{2\pi}\right)^{2}\right]\,,
\label{eq:tauafter}
\ea
where in the second equality we have used \eqref{eq:tauNf} and in the second line we have expanded in $\alpha/2\pi$, since $\alpha/2\pi<1$.
By defining $\Delta\tau\equiv\tau_{\{\alpha\}}-\tau$, we can read off the real and imaginary parts from \eqref{eq:tauafter} to give
\ba
{\rm Re}[\Delta\tau]&=&
\mathcal{A}\left\{-\left(c\frac{\theta}{2\pi}+d\right)^{2}+\left(c\frac{2\pi}{e^{2}}\right)^{2}\right\}\frac{\alpha}{2\pi}\cr\cr
&&+\,\mathcal{O}\left[\left(\frac{\alpha}{2\pi}\right)^{2}\right]\,,\cr\cr
{\rm Im}[\Delta\tau]
&=&\mathcal{A}\left\{-2\left(c\frac{\theta}{2\pi}+d\right)\left(c\frac{2\pi}{e^{2}}\right)\right\}\frac{\alpha}{2\pi}\cr\cr
&&+\,\mathcal{O}\left[\left(\frac{\alpha}{2\pi}\right)^{2}\right]\,.
\label{eq:Deltatau}
\ea

Now we are ready to infer what happens in the dual frame as a result of the $U(1)_{\rm chiral}$ transformation. After performing $U(1)_{\rm chiral}$, the functional (path) integral in the dual frame becomes 
\ba
&&\int\prod_{i=1}^{N_{f}}\mathcal{D}\Psi_{i}\, \mathcal{D}\bar{\Psi}_{i}\mathcal{D}A\,\, e^{iS-i\alpha\int{\rm d}\star j_{\rm chiral}}\cr\cr
&&=\int\prod_{i=1}^{N_{f}}\mathcal{D}\Psi_{i}\, \mathcal{D}\bar{\Psi}_{i}\mathcal{D}A\,\, e^{iS_{\{\alpha\}}}\,,
    \label{eq:functionalintegraltransf}
\ea
where $S$ is given in (\ref{eq:Sbefore}), $S_{\{\alpha\}}$ is (\ref{eq:Sbefore}) with $\tau$ replaced with $\tau_{\{\alpha\}}$ and $j_{\rm chiral}$ is the current generating $U(1)_{\rm chiral}$. By comparing the integrands in (\ref{eq:functionalintegraltransf}), we have
\ba
&&{\rm exp}\left[-i\alpha\int{\rm d}\star j_{\rm chiral}\right]\cr\cr
&&={\rm exp}\left[i\int-\frac{{\rm Im}[\Delta\tau]}{4\pi}F\wedge\star F+\frac{{\rm Re}[\Delta\tau]}{4\pi}F\wedge F\right]\,.\nonumber\\
\label{eq:expcompare}
\ea

In the limit of an infinitesimal chiral rotation, (\ref{eq:Deltatau}) and (\ref{eq:expcompare}) finally give us the ABJ anomaly equation in the dual frame
\ba
{\rm d}\star j_{\rm chiral}&=&-\frac{{\rm Re}[\Delta\tau]}{4\pi\alpha}F\wedge F+\frac{{\rm Im}[\Delta\tau]}{4\pi\alpha}F\wedge\star F\,,\cr\cr
&=&\frac{\mathcal{A}}{8\pi^{2}}\left\{\left(c\frac{\theta}{2\pi}+d\right)^{2}-\left(c\frac{2\pi}{e^{2}}\right)^{2}\right\}F\wedge F\cr\cr
&&-\frac{\mathcal{A}}{8\pi^{2}}\left\{2\left(c\frac{\theta}{2\pi}+d\right)\left(c\frac{2\pi}{e^{2}}\right)\right\}F\wedge\star F\,.\nonumber\\
\label{eq:ABJdyon2}
\ea
This result agrees with that obtained in Ref.\cite{Csaki:2010rv} where requiring the covariance of the equation of motion was invoked together with (\ref{eq:Ftransf}). In contrast, here we only relied on (\ref{eq:tauSL}) in tracking the variation of the action in the dual frame in parallel to that in the QED frame.

%%%%%%%%%%%%%%%%%%%%%%%%%%%%%%%%%%%%%%%%%%%%%%%%%%%%%%%%%%%%%%%%%%

\subsection{Non-invertible chiral symmetry in the dual frame}
\label{sec:INsymdyon}

The conventional understanding of the ABJ anomaly is that the classical, continuous $U(1)_{\rm chiral}$ symmetry is broken at the quantum level. But recently it was found that although non-invertible, a topological gauge-invariant symmetry operator can be constructed for every angle $\alpha=2\pi/k$ with a rational $k$~\cite{Choi:2022jqy,Cordova:2022ieu},
\ba
&&\mathcal{D}_{k}[{\rm d}A'/N]\cr\cr
&&=
\int[Da]\,{\rm exp}\left({\oint_{\Sigma}i\frac{2\pi p }{N}\star j_{\rm chiral}+\mathcal{A}^{N,p}[{\rm d}A'/N]}\right)\,,\nonumber\\
\label{eq:DpNelectron}
\ea
where $\Sigma$ is a closed and oriented 3-manifold, $N$ and $p$ are coprime integers such that $\mathcal{A}/k=p/N$. The Lagrangian, $\mathcal{A}^{N,p}[B^{[2]}]$ is the minimal $Z_{N}$ TQFT that couples to the two-form background gauge field $B^{[2]}$ of the one-form discrete global symmetry $Z_{N}^{[1]}$ and the 't Hooft anomaly of $Z_{N}^{[1]}$ is labeled by $p$~\cite{Hsin:2018vcg}. When $p=1$, the Lagrangian $\mathcal{A}^{N,1}[B^{[2]}]$ of the fractional quantum Hall state is
\be
\mathcal{A}^{N,1}\left[\frac{{\rm d}A'}{N}\right]=\frac{iN}{4\pi}a\wedge{\rm d}a+\frac{i}{2\pi}a\wedge {\rm d}A'\,.
\label{eq:ANpF}
\ee
Note that the background two-form gauge field $B^{[2]}$ of $Z_{N}^{[1]}$ on $\Sigma$ was identified with the dynamical two-form field in the bulk, and this gauges the $Z_{N}^{[1]}$ subgroup of $U(1)^{[1]}_{m}$ on the defect.

How the non-invertible symmetry defect $\mathcal{D}_{k}$ 
acts on local and line operators depends on whether the operator is affected by gauging the magnetic one-form symmetry. Since a fermion remains unaffected by gauging the magnetic one-form symmetry, $\mathcal{D}_{k}$ just acts on the fermion as an invertible, chiral rotation by the angle $\alpha=2\pi/k$. The existence of $\mathcal{D}_{k}$ is viewed as a legitimate chiral symmetry which imposes the same selection rules on correlators of local operators on $\mathbb{R}^{4}$ (or its IR regulated version $S^{4}$) as the invertible, discrete chiral rotation. The same is true for a Wilson line which remains unaffected by gauging the magnetic one-form symmetry, and thus $\mathcal{D}_{k}$ acts trivially. In contrast, since a 't Hooft line carries a magnetic charge, $\mathcal{D}_{k}$ acts on the 't Hooft line $T'_{q'_{\rm m}}(\gamma_{m})$ in the QED frame in a non-trivial way. When $T'_{q'_{\rm m}}(\gamma_{m})$ crosses the defect $\mathcal{D}_{k}$, the t'Hooft line $T'_{q'_{\rm m}}(\gamma_{m})$ becomes attached to the surface operator ${\rm exp}(i\frac{p}{N}\int F')$ to maintain gauge invariance.

Since the properties of the $U(1)$ gauge theory with matter fields should not depend on the specific choice of frame, we naturally expect that the non-invertible symmetry exists in the dual frame. To verify this expectation, note that the key features that allow for the construction of (\ref{eq:DpNelectron}) in massless QED with an anomalous $U(1)_{\rm chiral}$ symmetry were (i) there is a one-form magnetic symmetry with a conserved current $j_{m}^{\prime[2]}=\star F'/2\pi$ and (ii) the divergence of the anomalous $U(1)_{\rm chiral}$ symmetry current is given in the form 
\ba
{\rm d}\star j_{\rm chiral}&=&\frac{\mathcal{A}}{2}\,\frac{F'}{2\pi}\wedge\frac{F'}{2\pi}
=\frac{\mathcal{A}}{2}\star j_{m}^{\prime[2]}\wedge\star j_{m}^{\prime[2]}\,.
\label{eq:djchiralFF}
\ea
This implies that if (\ref{eq:ABJdyon2}) can be written as
\be
{\rm d}\star j_{\rm chiral}=\frac{\mathcal{A}}{2}
\star\hat{j}^{[2]}\wedge\star\hat{j}^{[2]}\,,
\label{eq:completesquare}
\ee
with a conserved current $\hat{j}^{[2]}$ in the dual frame, then this proves the presence of the non-invertible chiral symmetry in the dual frame. 
The non-invertible symmetry defect can then be simply constructed by setting $B^{[2]}=\frac{2\pi}{N}\star\hat{j}^{[2]}$ in $\mathcal{A}^{N,p}[B^{[2]}]$.

By comparing the expressions (\ref{eq:ABJdyon2}) and (\ref{eq:completesquare}), ${\rm d}\star j_{\rm chiral}$ in the dual frame can be written in terms of the following two-form 
\be
\star\hat{j}^{[2]}=\left(c\frac{\theta}{2\pi}+d\right)\frac{F}{2\pi}-c\frac{\star F}{e^{2}}\,,
\label{eq:Jhatdyon2}
\ee
up to the anomaly coefficient $\mathcal{A}$.
But then from the definition of the dual gauge field in (\ref{eq:dualA}) and its transformation rule in (\ref{eq:ASL}) under $SL(2,\mathbb{Z})$, we obtain
\ba
\frac{{\rm d}A'}{2\pi}&=&c\,\frac{{\rm d}\tilde{A}}{2\pi}+d\,\frac{{\rm d}A}{2\pi}\,,\cr\cr
&=&c\,\left(-\frac{1}{e^{2}}\star F+\frac{\theta}{2\pi}\frac{F}{2\pi}\right)+d\,\frac{F}{2\pi}\,,\cr\cr
&=&\star\hat{j}^{[2]}\,.
\label{eq:2formcurrent}
\ea
The identification in Eq.(\ref{eq:2formcurrent}) implies that ${\rm d}\star \hat{j}^{[2]}=0$, revealing that the  current (\ref{eq:Jhatdyon2}) in the dual frame is actually physically equivalent to the current generating the magnetic $U(1)^{[1]}_{m}$ symmetry  in the QED frame. Therefore, co-closedness of $\hat{j}^{[2]}$ is guaranteed by the Bianchi identity ${\rm d}F'=0$ as long as magnetic monopoles are absent in the QED frame. 

Analogously to the 't Hooft lines $T'_{q'_{\rm m}}(\gamma_{m})$  
in the QED frame, there must be dyon line operators in the dual frame on which the non-invertible symmetry defect $\mathcal{D}_{k}[\frac{2\pi}{N}\star\hat{j}^{[2]}]$ acts in a non-trivial manner. In other words, there is a line operator in the dual frame with a charge that is mutually non-local to the charge of the dyon field,
%that is the counterpart of a 
corresponding to the purely electrically-charged object in the QED frame.

As the charge $(0,q'_{m})$ of the magnetic monopole associated with $T'_{q'_{\rm m}}(\gamma_{m})$ in the QED frame is mapped to $(bq'_{m},aq'_{m})$ in the dual frame using (\ref{eq:QSL}), we can describe the dyon line operator of our interest in the dual frame as $D_{bq'_{m},aq'_{m}}(\gamma_{m})=W_{bq'_{m}}(\gamma_{m})T_{aq'_{m}}(\gamma_{m})$ which is the product of the Wilson line $W_{bq'_{m}}(\gamma_{m})$ and the 't Hooft line $T_{aq'_{m}}(\gamma_{m})$ in the dual frame. This corresponds to the worldline $\gamma_{m}$ of an infinitely heavy dyon charged under the one-form symmetry generated by $\hat{j}^{[2]}$ with the charge $\int_{\Sigma}\star\hat{j}^{[2]}=-c (b\,q'_{m})+d(a\,q'_{m})=q'_{m}$ 
in the dual frame.

Not surprisingly, this charge is invariant under $SL(2,\mathbb{Z})$ because $\int_{\Sigma}\star\hat{j}^{[2]}$ is actually equivalent to twice the pairwise helicity (\ref{eq:pairwise}) of an electron with $(Q'_{\rm e},Q'_{\rm m})=(1,0)$ and the heavy magnetic monopole with  $(Q'_{\rm e},Q'_{\rm m})=(0,q'_{m})$ in the QED frame. Similarly, in the dual frame, it is equivalent to twice the pairwise helicity of the dyon (counterpart of the electron in the QED frame) with $(Q_{\rm e},Q_{\rm m})=(d,c)$ and another heavy dyon (counterpart of the heavy magnetic monopole in the QED frame) with $(Q_{\rm e},Q_{\rm m})=(bq'_{m},aq'_{m})$.

Since $D_{bq'_{m},aq'_{m}}(\gamma_{m})$ 
transforms under the gauged $Z_{N}^{[1]}$ subgroup of the one-form symmetry generated by $\hat{j}^{[2]}$ in (\ref{eq:Jhatdyon2}) when it crosses the defect $\mathcal{D}_{k}[\frac{2\pi}{N}\star\hat{j}^{[2]}]$, it needs to be dressed by the surface operator ${\rm exp}(i\frac{2\pi p}{N}\int\star\hat{j}^{[2]})$ to maintain gauge invariance. Thus, under the non-invertible, discrete, chiral rotation with $\alpha=2\pi p/(\mathcal{A}N)$ in the dual frame, the dyon line operator transforms as
(see Appendix~\ref{sec:AppendixB})
\be
D_{bq'_{m},aq'_{m}}(\gamma_{m})\,\,\rightarrow\,\,D_{bq'_{m},aq'_{m}}(\gamma_{m})e^{i\frac{p}{N}\int2\pi\star\hat{j}^{[2]}}\,.
\label{eq:dyonlinetransf}
\ee
This is the analogue of the transformation of the 't Hooft line operator in the QED frame under the non-invertible discrete chiral rotation with $\alpha=2\pi p/(\mathcal{A}N)$ 
\be
T'_{q'_{\rm m}}(\gamma_{m})\,\,\rightarrow\,\, T'_{q'_{\rm m}}(\gamma_{m})e^{i\frac{p}{N}\int F'}\,.
\label{eq:tHooftlinetransf}
\ee
%The transformation of the dyon line operator  (\ref{eq:dyonlinetransf}) is explicitly shown in Appendix~\ref{sec:AppendixB}.

%%%%%%%%%%%%%%%%%%%%%%%%%%%%%%%%%%%%%%%%%%%%%%%%%%%%%%%%%%%%%%%%%%

\section{Implications}
In this section, we present two implications of the non-invertible chiral symmetry in the dual frame. We first discuss the fact that the non-invertible chiral symmetry is necessary to consistently understand the role of the kinetic and topological terms in any frame. Second, we show that in axion electrodynamics, the presence of a non-invertible symmetry dictates whether an axion potential can be radiatively generated.

\label{sec:implications}
\subsection{Chiral rotation at a fixed point}

The form of the anomaly equation in the dual frame \eqref{eq:ABJdyon2} suggests the kinetic term $F\wedge \star F$ may be rotated away by a chiral transformation, implying the dynamics can be altered under such a transformation. To explicitly check that the equation of motion does not change in the dual frame following a chiral transformation, we consider the theory at a specific infrared fixed point, for simplicity.

Consider an $SL(2,\mathbb{Z})$ transformation with $c=Q_{\rm m}/Q'_{\rm e}$ and $d=Q_{\rm e}/Q'_{\rm e}$, that maps the charge $(Q'_{\rm e},0)$ in the 
QED frame to $(Q_{\rm e},Q_{\rm m})$ in the dual frame. Using the result (\ref{eq:betadyon}), we immediately see that gauge coupling $\beta$-function, $\beta_{e}$, vanishes provided
\be
 Q_{\rm m}\frac{\theta}{2\pi}+Q_{\rm e}=\pm Q_{\rm m}\frac{2\pi}{e^{2}}\,.
\label{eq:conditionzerobeta}
\ee
Note that when (\ref{eq:conditionzerobeta}) is satisfied, the contributions to the photon self energy from electric and magnetic charges are the same in the magnitude. Since their contributions have opposite sign~\cite{Laperashvili:1999pu}, the vanishing $\beta$-function, $\beta_{e}=0$, can be understood as resulting from the cancellation between these two contributions.

Assuming (\ref{eq:conditionzerobeta}) holds, one can analytically solve the renormalization group equation of $\theta$ in the dual frame using the expression (\ref{eq:betadyon}) to obtain
\be
\theta(t)=\mp\frac{2\pi\left[2\pi Q^{'2}_{\rm e}\pm\left(\frac{16\pi^{2}}{e'^{3}}\beta_{e'}\right)Q_{\rm e}Q_{\rm m}t+4\pi^{2}Q_{\rm e}Q_{\rm m}Q^{'2}_{\rm e}c_{\theta}\right]}{\left[\left(\frac{16\pi^{2}}{e'^{3}}\beta_{e'}\right)t\pm4\pi^{2}Q^{'2}_{\rm e}c_{\theta}\right]Q_{\rm m}^{2}}\,,
\label{eq:thetat}
\ee
where $t=\ln(\mu/\mu_{0})$ with $\mu_{0}$ an infrared (IR) renormalization scale and $c_{\theta}$ is an integration constant. For $t=0$ and a sufficiently large $c_{\theta}$, we see from (\ref{eq:thetat}) that $\beta_{\theta}=0$ is realized in the IR at $\theta=-2\pi Q_{\rm e}/Q_{\rm m}$. As a matter of fact, we could have reached the same conclusion by simply solving
\ba
\pm\frac{16\pi^{2}}{e'^{3}}\beta_{e'}\left(c\frac{\theta}{2\pi}+d\right)^{2}=0\,,
\ea
which is equivalent to $\beta_{\theta}=0$ when (\ref{eq:conditionzerobeta}) holds. Finally, by substituting $\theta=-2\pi Q_{\rm e}/Q_{\rm m}$ into (\ref{eq:conditionzerobeta}), we find the IR fixed point (IRFP) of the dual frame 
\be
{\bf{ IRFP}}:\quad(\theta_{*},e_{*})=(-2\pi\frac{Q_{\rm e}}{Q_{\rm m}},\infty)\,.
\label{eq:IRFP}
\ee
The fixed point can also be written as
\be
\tau_{*}=\frac{\theta_{*}}{2\pi}+i\frac{2\pi}{e_{*}^{2}}=-\frac{Q_{\rm e}}{Q_{\rm m}}+i\epsilon_{*}\,,
\label{eq:taustar}
\ee
where $\epsilon_{*}\equiv2\pi/e_{*}^{2}\ll1$. Applying $SL(2,\mathbb{Z})$ to (\ref{eq:taustar}), we find that the IRFP of the dual frame in (\ref{eq:taustar}) corresponds to $e'=\sqrt{2\pi\epsilon_{*}(Q_{\rm m}/Q'_{\rm e})^{2}}\rightarrow 0$ in the QED frame which is the well-known IR fixed point of QED.

Intuitively, this interesting feature of the theory in the dual frame should not be affected by the $U(1)_{\rm chiral}$ rotation with the non-invertible $U(1)_{\rm chiral}$. At least, this point is clear in the QED frame as the full $U(1)_{\rm chiral}$ rotation does not affect the IR fixed point of QED. However, because at the IRFP in the dual frame, (\ref{eq:ABJdyon2}) suggests that
\be
{\rm d}\star j_{\rm chiral}=-\frac{\mathcal{A}}{8\pi^{2}}\left\{2\left(c\frac{\theta}{2\pi}+d\right)\left(c\frac{2\pi}{e^{2}}\right)\right\}F\wedge\star F\,,
\label{eq:ABJdyon3}
\ee
it appears to be possible to %approximately 
rotate away the $-\frac{1}{2e^{2}}F\wedge\star F$ kinetic term in the Lagrangian by a $U(1)_{\rm chiral}$ rotation. At first glance, this would imply that the action only contains the $\frac{\theta}{8\pi^{2}}F\wedge F$ term, contradicting the fact that the gauge coupling is renormalized with $\beta_{e}=0$.

To resolve this apparent contradiction, we crucially note that $F\wedge F$ is no longer a total derivative in the presence of dyons, and thus varying the action with respect to $A$ we obtain the equation of motion
\be
{\rm d}\left(-\frac{\theta_{*}}{2\pi}\frac{F}{2\pi}\right)=\star J\,.
\label{eq:currentelectricFP}
\ee
On the other hand, since the removal of the $-\frac{1}{2e^{2}}F\wedge\star F$ term can be performed with the full $U(1)_{\rm chiral}$ containing the unbroken non-invertible subgroup, the one-form symmetry in the dual frame generated by the two-form current in (\ref{eq:Jhatdyon2}) should remain unbroken, i.e. ${\rm d}\star\hat{j}^{[2]}=0$. When applied to (\ref{eq:Jhatdyon2}), this means that at the IRFP characterized by (\ref{eq:conditionzerobeta}), we actually have the non-trivial relation
\be
{\rm d}F=\pm{\rm d}\star F\,.
\label{eq:dFequaldstarF}
\ee
Therefore, applying (\ref{eq:dFequaldstarF}) to (\ref{eq:currentelectricFP}), we obtain
\be
{\rm d}\left(\mp\frac{\theta_{*}}{2\pi}\frac{\star F}{2\pi}\right)=\star J\,,
\label{eq:currentelectric2}
\ee
where $\theta_{*}$ is related to the gauge coupling fixed point $e_{*}$ via (\ref{eq:conditionzerobeta}). This is the equation of motion obtained at the fixed point after rotating away the $F\wedge \star F$ term.

Contrary to the naive guess that the kinetic term can be rotated away, remarkably (\ref{eq:currentelectric2}), resulting from ${\rm d}\star\hat{j}^{[2]}=0$, shows that a kinetic term 
is still present, even after implementing the full $U(1)_{\rm chiral}$ to remove the $F\wedge \star F$ term. This explicitly reveals that each of the kinetic and topological terms are all mixtures of $F\wedge F$ and $F\wedge\star F$ in the dual frame. Indeed, just starting from the action in the dual frame at the IRFP (\ref{eq:IRFP}), namely 
\be
S\supset \int-\frac{1}{2e_{*}^{2}}F\wedge\star F+\frac{\theta_{*}}{8\pi^{2}}F\wedge F-A\wedge\star J\,,
\ee
before using $U(1)_{\rm chiral}$ to remove the $-\frac{1}{2e^{2}}F\wedge\star F$ term, one obtains the equation of motion of $A$
\ba
\star J&=&{\rm d}\left(\frac{1}{e_{*}^{2}}\star F-\frac{\theta_{*}}{2\pi}\frac{F}{2\pi}\right)\,,\cr\cr
&=&{\rm d}\left(\pm\frac{\theta_{*}}{2\pi}\frac{\star F}{2\pi}\pm\frac{Q_{\rm e}}{Q_{\rm m}}\frac{\star F}{2\pi}-\frac{\theta_{*}}{2\pi}\frac{F}{2\pi}\right)\,,\cr\cr
&=&{\rm d}\left(\mp\frac{\theta_{*}}{2\pi}\frac{\star F}{2\pi}\right)\,,
\label{eq:currentelectric3}
\ea
where 
(\ref{eq:conditionzerobeta}) was used to replace $e_{*}^{-2}$ in the second line, and (\ref{eq:dFequaldstarF}),(\ref{eq:IRFP}) were used to obtain the third line. Hence, the agreement between (\ref{eq:currentelectric2}) and (\ref{eq:currentelectric3}) directly proves that the non-invertible $U(1)_{\rm chiral}$ symmetry has no effect on the equation of motion of $A$ as is the case in the QED frame. 

Of course, this property is not exclusive to the IRFP; it also manifests at any point along the renormalization group flow as a generic feature. Provided either of $F\wedge\star F$ or $F\wedge F$ were rotated away by means of the non-invertible $U(1)_{\rm chiral}$ and (\ref{eq:ABJdyon2}), the fact that ${\rm d}\star\hat{j}^{[2]}=0$ allows the remaining term to still produce the correct equation of motion. On the other hand, when ${\rm d}\star\hat{j}^{[2]}\neq0$ and the pairwise helicity is nonzero, the non-invertible subgroup of $U(1)_{\rm chiral}$ is broken and thus we are no longer able to freely rotate away $F\wedge\star F$ or $F\wedge F$ in the Lagrangian because the classical $U(1)_{\rm chiral}$ no longer exists.

\subsection{Axion electrodynamics in the dual frame}

The prevailing understanding for why there is no axion potential in the axion-Maxwell theory has largely been based on the supposed absence of Abelian instantons in $S^{4}$. Since Abelian instantons have been explicitly constructed \cite{Csaki:2024ajo,GarciaGarcia:2025uub} in theories with both electric and magnetic charges and non-trival spacetime topologies, we need a more rigorous explanation. We will now show that this phenomenon can be understood through the selection rule imposed by the non-invertible chiral symmetry. In particular, we will see that the existence of non-zero pairwise helicities can indicate whether there is
a non-vanishing axion potential.

This new perspective is very intriguing and of practical relevance, as it sheds light on a concrete way to generate an axion potential. The violation of the selection rule from the non-invertible chiral symmetry by a non-vanishing axion potential indicates a strong connection between the conditions required for the breaking of the non-invertible chiral symmetry and generating an axion potential. Thus, the symmetry perspective naturally leads us to the idea that the breaking of the one-form symmetry, i.e. $U(1)_{m}^{[1]}$ in the QED frame and its counterpart in the dual frame, by fields with charges that are mutually non-local to the charges in the low-energy theory is strongly tied with generating an axion potential. This symmetry perspective is consistent with the axion mass generation via magnetic monopole loops~\cite{Fan:2021ntg,Cordova:2022ieu} in a theory with a light electron, and with \cite{GarciaGarcia:2025uub} which showed how Abelian instanton configurations are associated with dyon loops.

In the dual frame, introducing fields with mutually non-local charges breaks the one-form symmetry generated by $\hat{j}^{[2]}$ given in (\ref{eq:Jhatdyon2}), causing the breaking of the non-invertible $U(1)_{\rm chiral}$. To explicitly characterize how this occurs, we can study how (\ref{eq:unclosedF}) in the QED frame transforms under $SL(2,\mathbb{Z})$. 
Alternatively, we can apply the $SL(2,\mathbb{Z})$ transformation to the imaginary part of (\ref{eq:combinedMW}) in the QED frame. This leads to
\ba
{\rm d}\star\hat{j}^{[2]}&=&\frac{{\rm Im}[(c\tau^{*}+d)(\star J+\tau\star K)]}{{\rm Im}[\tau]}\,,\cr\cr
&=&-c\,\star J+d\,\star K\,,
\label{eq:d2formcurrent}
\ea
where $J$ and $K$ are the electric and magnetic currents in the dual frame, respectively, and $\star\hat{j}^{[2]}$ in (\ref{eq:d2formcurrent}) is written as
\be
\star\hat{j}^{[2]}={\rm Im}[(c\tau^{*}+d)({\rm d}\star F+i{\rm d}F)]\,.
\label{eq:Jhatdyon3}
\ee
Note that (\ref{eq:Jhatdyon3}) is equivalent to $\star\hat{j}^{[2]}$ in (\ref{eq:Jhatdyon2}).

In the dual frame, it is important to note that the ratio of electric and magnetic charges are constrained in order to preserve a non-invertible chiral symmetry. To see this, let us first introduce a reference dyon in the dual frame, and then choose $(c,d)$ as in (\ref{dualityparam}) so that the reference dyon in the dual frame is mapped to an electrically charged particle in the QED frame. Now when a second dyon is introduced, in addition to the reference dyon, as long as the pairwise helicity, (\ref{eq:pairwise}), between 
the reference dyon and the new dyon is zero, the non-invertible chiral symmetry is preserved.  More explicitly, 
denoting the charge of the second dyon as $(Q_{\rm e2}, Q_{\rm m2})$, we obtain a new term on the right-hand side of (\ref{eq:d2formcurrent}), given by
\be
 -c\, Q_{\rm e2}+ d\, Q_{\rm m2}=  -\frac{1}{n}\left(Q_{\rm m} Q_{\rm e2}- Q_{\rm e} Q_{\rm m2}\right)~,
\ee
which is just proportional to the pairwise helicity \eqref{eq:pairwise}. This implies that when there are fields in any frame with non-vanishing pairwise helicity, the one-form symmetry generated by $\hat{j}^{[2]}$ is broken and the non-invertible chiral symmetry no longer exists. In this way, we see that the pairwise helicity acts as a ``spurion'' for the non-invertible symmetry.

In the case where the one-form symmetry is broken, we may ask if the theory can admit a non-vanishing axion potential.
To this end, we can check if a non-zero topological susceptibility, $\chi$ is allowed without any inconsistency. The definition of $\chi$ is
\be
\chi=\frac{\delta^{2}\log\,Z(\theta)}{\delta\theta(x)\theta(0)}=\int d^4x\, \langle q(x) q(0) \rangle\,,
\label{eq:TS}
\ee
where $\theta(x)\equiv a(x)/f$ is the ratio of the axion field $a(x)$ to the axion decay constant $f$, $Z(\theta)$ is the partition function, and the topological charge density in the QED frame is
\be
q(x) = \frac{1}{16\pi^2} F'_{\mu\nu}(x) \tilde{F}^{\prime\mu\nu}(x)\,.
\ee
The integral of $q(x)$ over a four-dimensional (spin) manifold $M$ gives the topological charge $Q_T$ in the QED frame 
\be
Q_T=\int_{M}\, q(x)=\frac{1}{8\pi^{2}}\int_{M}F'\wedge F'\,.
\label{eq:topologicalcharge}
\ee 
Note that because $j^{\prime[2]}_{m}=\frac{\star F'}{2\pi}$ transforms under $SL(2,\mathbb{Z})$, the topological charge in the dual frame becomes
\be
Q_T=\frac{1}{2}\int_{M}\star\hat{j}^{[2]}\wedge\star\hat{j}^{[2]}\,,
\label{eq:qxjj}
\ee
where $\star\hat{j}^{[2]}$ is given in (\ref{eq:Jhatdyon2}).

When the one-form magnetic symmetry is broken due to mutually non-local charges, $Q_T$ takes integer values due to a non-trivial spacetime topology that includes two-cycles or an Abelian instanton~\cite{Csaki:2024ajo,GarciaGarcia:2025uub}. Under the chiral rotation by an angle $\alpha$, the fermion measure of the path integral then transforms as~\cite{Fujikawa:1979ay,Fujikawa:1980eg} 
\ba
\prod_{j=1}^{N_{f}}\mathcal{D}\Psi_{j}\, \mathcal{D}\bar{\Psi}_{j} \,\;\rightarrow\prod_{j=1}^{N_{f}}\mathcal{D}\Psi_{j}\, \mathcal{D}\bar{\Psi}_{j}\,\, e^{-i\alpha\mathcal{A} Q_T}\,.
    \label{eq:measuretransf}
\ea
From (\ref{eq:measuretransf}), we see that when $\alpha$ is an integral multiple of $2\pi/\mathcal{A}$, the functional integral remains invariant since the charge $Q_T\in \mathbb{Z}$. Hence, when the non-invertible symmetry is broken by mutually non-local charges, there is residual, invertible, discrete subgroup $Z_{\mathcal{A}}\subset U(1)_{\rm chiral}$. A non-vanishing axion potential (equivalently, $\chi\neq0$) is then allowed as long as $a(x)\sim a(x)+2\pi f/\mathcal{A}$ is a symmetry of the potential~\cite{Cordova:2022ieu}. In \cite{Fan:2021ntg}, the axion mass induced by the magnetic monopole loop in the QED frame was computed by applying the worldline formalism to the dyonic mode on a magnetic monopole worldline. The nonzero axion mass was expected since the magnetic monopole is mutually non-local to an electrically charged fermion.

One may actually consider QCD axion scenarios with mutually non-local fermions (i.e. non-zero pairwise helicity). 
For instance, a variant of the KSVZ model~\cite{Kim:1979if,Shifman:1979if} was considered in \cite{Sokolov:2022fvs,Sokolov:2023pos}, where a heavy vector-like dyonic fermion charged under the anomalous chiral symmetry (or PQ symmetry) was introduced in addition to the electrically-charged, PQ-neutral, Standard Model leptons. Since the pairwise helicity between the dyonic fermion and the SM electron (or other SM leptons) is always non-zero, we expect ${\rm d}\star\hat{j}^{[2]}\neq 0$ (or ${\rm d}\star j_{m}^{\prime[2]}\neq 0$) in any $U(1)_{\rm EM}$ frame.
Thus, in such models, since the non-invertible symmetry is explicitly broken, the axion can obtain a new contribution to its mass in addition to the usual QCD contribution~\cite{Heidenreich:2023pbi}.

It is interesting to consider the opposite case, where only dyons that are mutually local (i.e. zero pairwise helicity)
are present in the dual frame with axion electrodynamics. By promoting $\alpha$ to a periodic dynamical axion field $a(x)/f\in[0,2\pi]$ in Sec.~\ref{sec:ABJanomaly}, the axion-photon coupling in the dual frame is obtained from \eqref{eq:ABJdyon2}, namely
\ba
S&\supset&\int\mathcal{A}\left\{\left(c\frac{\theta}{2\pi}+d\right)^{2}-\left(c\frac{2\pi}{e^{2}}\right)^{2}\right\}\frac{a(x)}{8\pi^{2}f}F\wedge F\cr\cr
&&-\int\mathcal{A}\left\{2\left(c\frac{\theta}{2\pi}+d\right)\left(c\frac{2\pi}{e^{2}}\right)\right\}\frac{a(x)}{8\pi^{2}f}F\wedge\star F\,.\nonumber\\
\label{eq:axionphoton}
\ea
Thus, (\ref{eq:axionphoton}) is the axion-photon coupling when fermions contributing to the coupling are mutually local to each other.
But when they are mutually non-local, the coupling has non-perturbative contributions~\cite{Csaki:2024plt,Csaki:work}.
Curiously, one sees that $a(x)$ not only couples to $F\wedge F$, but also $F\wedge \star F$ \cite{Sokolov:2022fvs}. The axion coupling to $F\wedge \star F$, suggests that it might be possible to radiatively generate an axion potential in the dual frame leading to different physics in the two frames related by $SL(2,\mathbb{Z})$ duality. However, this cannot be the case since having mutually local dyons  implies ${\rm d}\star\hat{j}^{[2]}=0$ and the non-invertible chiral symmetry remains unbroken. Hence, we expect the radiative generation of an axion potential in the dual frame with mutually local charges is in fact forbidden by the non-invertible chiral symmetry as is manifest in the QED frame with only massless electrons. Since the pairwise helicity is zero, one can also transform to the QED frame where the dyons contributing to (\ref{eq:axionphoton}) become electrons. Then, the axion couples to only $F'\wedge F'$ in the QED frame and thus the absence of the radiatively induced axion potential becomes clearer.

Before we prove this, let us first discuss the fate of the invertible discrete subgroup $Z_{\mathcal{A}}$. When there are only mutually local charges in the QED frame, the topological charge $Q_T$ is still an integer because $F$ represents a class in the integer cohomology of degree two, $H^{2}(M,\mathbb{Z})$ and thus $F\wedge F$ represents a class in the integer cohomology of degree four, $ H^{4}(M,\mathbb{Z})$~\cite{Bott:1982xhp,Nakahara:2003nw}. Similarly, in the dual frame we have $2\pi\star\hat{j}^{[2]}\in H^{2}(M,\mathbb{Z})$ and thus $4\pi^{2}\star\hat{j}^{[2]}\wedge\star\hat{j}^{[2]}\in H^{4}(M,\mathbb{Z})$. Since we are working on a spin manifold $M$, we obtain
\be
\frac{1}{8\pi^{2}}\int_{M}F'\wedge F'=\frac{1}{2}\int_{M}\star\hat{j}^{[2]}\wedge\star\hat{j}^{[2]}\in\mathbb{Z}\,,
\label{eq:integralFFJJ}
\ee
implying that $Q_{T}$ is integrally quantized. Alternatively, (\ref{eq:integralFFJJ}) can also be understood from the index theorem where $\int_{M}{\rm d}\star j_{\rm chiral}=n_+ - n_-$, with $n_{\pm}$ are the number of right (left)-handed fermions
and \eqref{eq:completesquare} with ${\cal A}\in \mathbb{Z}$ has been used. Thus, even when ${\rm d}\star\hat{j}^{[2]}=0$, there is a valid invertible discrete subgroup $Z_{\mathcal{A}}$.

To prove that $\chi=0$ is enforced by the selection rule of the non-invertible chiral symmetry, one can show that $\delta_{\alpha}\chi=0$ under the action of $\mathcal{D}_{k}$ is possible only when $\chi=0$.  To compute the variation $\delta_{\alpha}\chi$, recall that under an infinitesimal transformation $\alpha$, the non-invertible chiral symmetry operator $\mathcal{D}_{k}$ acts on fermion fields $\Psi_{j}$ as the invertible chiral rotation with $\alpha=2\pi/k=2\pi p/(\mathcal{A}N)$. Thus from (\ref{eq:measuretransf}), the expectation value of a gauge-invariant local operator $\mathcal{O}$ transforms as 
\ba
\delta_{\alpha}\langle\mathcal{O}\rangle&=&-i\alpha\mathcal{A}\langle Q_{T}\,\mathcal{O}\rangle\,,\cr\cr
&=&-i\alpha\mathcal{A}\int d^{4}y\,\langle q(y)\mathcal{O}\rangle\,
.
\label{eq:variationO}
\ea
Note that since the path integral sums over all possible gauge field configurations in $Q_T$, the topological charge $Q_T$ cannot be simply factored out of the angle brackets in (\ref{eq:variationO}). Now, when (\ref{eq:variationO}) is applied to the topological susceptibility in (\ref{eq:TS}) we obtain
\ba
    \delta_\alpha \chi &=& \int d^4x\, \delta_\alpha\langle q(x) q(0)\rangle\,,\cr\cr
    &=&
    -i\alpha\mathcal{A} \int d^4x\, \langle Q_T\,q(x)\, q(0) \rangle \,,\cr\cr
    &=&-i \alpha\mathcal{A} \int d^4x\, d^4y\, \langle q(y) q(x) q(0) \rangle\,,
\label{eq:deltachi}
\ea
where (\ref{eq:topologicalcharge}) was used in the second line. 

Note that when the non-invertible chiral symmetry is present, it means $\delta_{\alpha}\chi=0$ is required for every angle $\alpha=2\pi/k=2\pi p/(\mathcal{A}N)$ with coprime $p$ and $N$, thanks to the topological and gauge invariant defect in (\ref{eq:DpNelectron}) in the QED frame (or $\mathcal{D}_{k}$ with $\mathcal{A}^{N,p}[\star\hat{j}^{[2]}/N]$ in the dual frame). This condition cannot be satisfied for an arbitrary field configuration with $\chi\neq 0$ because the three-point correlator of the topological charge density \eqref{eq:deltachi} would be non-zero in general. Thus, the selection rule $\delta_{\alpha}\chi=0$ imposed by the non-invertible symmetry is satisfied if and only if $\chi=0$. This proves that an axion potential cannot be generated when there exists a non-invertible chiral symmetry in the theory.

%%%%%%%%%%%%%%%%%%%%%%%%%%%%%%%%%%%%%%%%%%%%%%%%%%%%%%%%%%%%%%%%%%

\section{Conclusion}
In this paper we have constructed the non-invertible chiral symmetry in a $U(1)$ gauge theory with massless fermionic dyons, thereby confirming its presence in the $SL(2,\mathbb{Z})$ dual frame of massless QED. In Sec.~\ref{sec:ABJanomaly}, we first derived the ABJ anomaly equation in the dual frame based on the combined action of the $SL(2,\mathbb{Z})$ transformation and $U(1)_{\rm chiral}$ rotation on the complexified gauge coupling $\tau.$ Armed with this, in Sec.~\ref{sec:INsymdyon} we identified a conserved current $\hat{j}^{[2]}$ that generates a one-form symmetry in the dual frame, corresponding to the magnetic one-form symmetry in the QED frame. This enabled us to show that the chiral anomaly in the dual frame is given in the form ${\rm d}\star j_{\rm chiral}\propto\star \hat{j}^{[2]}\wedge\star \hat{j}^{[2]}$ and hence a non-invertible chiral symmetry is present in all duality frames, as expected.

We also explored several implications of this construction in Sec.~\ref{sec:implications}. In the dual frame, we argued that the non-invertible chiral symmetry is necessary to maintain a manifest frame-independent description of the physics. This was explicitly shown at a fixed point where although the ABJ anomaly (\ref{eq:ABJdyon3}) appears to indicate that the gauge kinetic term can be eliminated by the full $U(1)_{\rm chiral}$ rotation, the  $\theta$-term still contains the gauge kinetic term. This is consistent with the fact that the $U(1)_{\rm chiral}$ rotation does not affect the renormalization group flow of the theory in the QED frame.

In the context of axion electrodynamics, whenever there are mutually non-local dyons, thus violating ${\rm d}\star\hat{j}^{[2]}=0$ in the dual frame, $U(1)_{\rm chiral}$ breaks down to the invertible subgroup $Z_{\mathcal{A}}$. We proved that a non-zero topological susceptibility and thus an axion potential invariant under $a(x)\sim a(x)+2\pi f/\mathcal{A}$ is allowed. In contrast, when ${\rm d}\star\hat{j}^{[2]}=0$ holds in the dual frame, the non-invertible chiral symmetry $\mathcal{D}_{k}$ remains a good symmetry of the theory and we proved that the only way for $\chi$ to obey the selection rule imposed by $\mathcal{D}_{k}$ is $\chi=0$, implying no axion potential is generated.

Our analysis demonstrates the compatibility of non-invertible symmetries with duality transformations and reinforces their role as intrinsic features of quantum field theories. Also we have clarified the role of the pairwise helicity as a ``spurion" for the breaking of non-invertible chiral symmetry. It would be interesting to investigate a UV completion for the case with a non-zero pairwise helicity as well as apply our methods to similar constructions in more general theories, such as $U(1)\times U(1)$ describing photons mixing with dark photons which, with the existence of mutually non-local charges becomes nontrivial \cite{Terning:2018lsv}, and could lead to new phenomenological applications.

\medskip\noindent\textit{Acknowledgments\,---\,}%
We thank S. Chen, A. Cherman, C. Cordova, C.~Cs{\'a}ki, E. Ievlev, J. Newey, R.~Ovadia, M.~Ruhdorfer, O.~Telem, and C. Verhaaren for helpful discussions. We are also grateful to A. Cherman and S.-H. Shao for comments on the manuscript. The work of G.C. and T.G. is supported in part by the Department of Energy under grant DE-SC0011842 at the University of Minnesota. The work of J.T. is supported in part by the Department of Energy under grant DE-SC0009999. T.G. and J.T. thank the Aspen Center for Physics, which is supported by National Science Foundation grant PHY-2210452 where this work was initiated, and also acknowledge the Kavli Institute for Theoretical Physics (KITP) in Santa Barbara, supported by grant NSF PHY-2309135, and the Munich Institute for Astro-, Particle and BioPhysics (MIAPbP), funded by the Deutsche Forschungsgemeinschaft (DFG, German Research Foundation) under Germany's Excellence Strategy – EXC-2094 – 390783311, where portions of this work were done.

\appendix

\section{Running of $\tau$ in the dual frame}
\label{sec:AppendixA}

In this appendix, the $\beta$-function of the complexified gauge coupling in the dual frame is derived by applying the $SL(2,\mathbb{Z})$ transformation to the $\beta$-function in the QED frame.

In the QED frame, using the known $\beta$-function in QED, we can write down the $\beta$-function of the complexified gauge coupling as
\be
\frac{{\rm d}\tau'}{{\rm d}\ln\mu}=-\frac{4\pi}{e'^{3}}\frac{{\rm d}e'}{{\rm d}\ln\mu}i\,,
\label{eq:betataup}
\ee
where the $\beta$-function for $e'$ is given as
\be
\beta_{e'}=\frac{{\rm d}e'}{{\rm d}\ln\mu}=\sum_{\ell=1}^{\infty}e'\left[b'_{\ell}\left(\frac{e'^{2}}{16\pi^{2}}\right)^{\ell}\right]\,.
\ee
and the sum is over the number of loops.
For instance, the one-loop $\beta$-function coefficient for $N_f$ Dirac fermions is
\be
b'_{1}=\sum_{i=1}^{N_{f}}\frac{4}{3}Q_{e,i}^{'2}\,.
\ee

On the other hand, using (\ref{eq:tauSL}) with $ad-bc=1$, we can rewrite (\ref{eq:betataup}) as
\be
\frac{{\rm d}\tau}{{\rm d}\ln\mu}=-(c\tau+d)^{2}\frac{4\pi}{e'^{3}}\frac{{\rm d}e'}{{\rm d}\ln\mu}i\,.
\label{eq:betataudyon}
\ee
Then, from identifying the real and imaginary parts of the left and right hand sides of (\ref{eq:betataudyon}), we find
\ba
&&\beta_{\theta}\equiv\frac{{\rm d}\theta}{{\rm d}\ln\mu}=\frac{16\pi^{2}}{e'^{3}}\beta_{e'}\left(c\frac{\theta}{2\pi}+d\right)\left(c\frac{2\pi}{e^{2}}\right)\,,\cr\cr
&&\beta_{e}\equiv\frac{{\rm d}e}{{\rm d}\ln\mu}=\frac{e^{3}}{e'^{3}}\beta_{e'}\left\{\left(c\frac{\theta}{2\pi}+d\right)^{2}-\left(c\frac{2\pi}{e^{2}}\right)^{2}\right\}\,.\nonumber\\
\label{eq:betadyon}
\ea

Notice that the coefficients of $F\wedge F$ and $F\wedge\star F$ in (\ref{eq:ABJdyon2}) are proportional to $\beta_{e}$ and $\beta_{\theta}$ respectively. Therefore, we can rewrite the ABJ anomaly equation in the dual frame as
\be
{\rm d}\star j_{\rm chiral}=\frac{\mathcal{A}}{8\pi^{2}}\frac{e'^{3}\beta_{e}}{e^{3}\beta_{e'}}F\wedge F+\frac{\mathcal{A}}{8\pi^{2}}\frac{e'^{3}\beta_{\theta}}{8\pi^{2}\beta_{e'}}F\wedge\star F\,.\nonumber\\
\label{eq:ABJdyon4}
\ee

\section{The transformation of $D_{bq'_{m},aq'_{m}}(\gamma_{m})$ under the action of $\mathcal{D}_{k}$}
\label{sec:AppendixB}

The transformation of the 't Hooft line across the domain wall (i.e. the supporting manifold of $\mathcal{D}_{k}$) given in (\ref{eq:tHooftlinetransf}) can be viewed as the Witten effect since the $\theta$ parameter changes by $2\pi\mathcal{A}/k=2\pi p/N$. Thus, a similar effect will apply to the dyon line operator $D_{bq'_{m},aq'_{m}}(\gamma_{m})$ when it crosses the domain wall.

Analogous to the 't Hooft line which acquires the fractional electric charge $p/N$, the dyon line operator $D_{bq'_{m},aq'_{m}}(\gamma_{m})$ 
acquires fractional electric and magnetic charges, $d\,p/N$ and $c\,p/N$, respectively. Thus, when crossing the domain wall, the line operator $D_{bq'_{m},aq'_{m}}(\gamma_{m})$ becomes dressed by $W_{dp/N}(\gamma_{m})T_{cp/N}(\gamma_{m})$.

Using Stoke's theorem, we can then rewrite $W_{dp/N}(\gamma_{m})T_{cp/N}(\gamma_{m})$ as follows
\ba
&&W_{dp/N}(\gamma_{m})T_{cp/N}(\gamma_{m})\nonumber\\
&&={\rm exp}\left[i\frac{p}{N}\int_{\gamma_{m}} d\,A+c\,\tilde{A}\right]\,,\nonumber\\
&&={\rm exp}\left[i\frac{p}{N}\int_{\Sigma_{m}} \left(c\frac{\theta}{2\pi}+d\right)F-c\frac{2\pi}{e^{2}}\star F\right]\,,\nonumber\\
&&={\rm exp}\left[i\frac{p}{N}\int_{\Sigma_{m}}2\pi\star\hat{j}^{[2]}\right]\,,
\ea
where $\gamma_{m}=\partial\Sigma_{m}$ and (\ref{eq:Jhatdyon2}) was used to obtain the last line. This explains the dyon line operator transformation (\ref{eq:dyonlinetransf}) under the non-invertible chiral symmetry.

%%%%%%%%%%%%%%%%%%%%%%%%%%%%%%%%%%%%%%%%%%%%%%%%%%%%%%%%%%%%%%%%%%%%%%%%%%%%%%%%%%%%%%%%%%%%%%%%%%%%

\bibliographystyle{JHEP}
\bibliography{arxiv_1}

\providecommand{\href}[2]{#2}\begingroup\raggedright\begin{thebibliography}{10}

\bibitem{Gaiotto:2014kfa}
D.~Gaiotto, A.~Kapustin, N.~Seiberg, and B.~Willett, {\it {Generalized Global Symmetries}},  {\em JHEP} {\bf 02} (2015) 172, [\href{http://arxiv.org/abs/1412.5148}{{\tt arXiv:1412.5148}}].

\bibitem{Aharony:2013hda}
O.~Aharony, N.~Seiberg, and Y.~Tachikawa, {\it {Reading between the lines of four-dimensional gauge theories}},  {\em JHEP} {\bf 08} (2013) 115, [\href{http://arxiv.org/abs/1305.0318}{{\tt arXiv:1305.0318}}].

\bibitem{Gomes:2023ahz}
P.~R.~S. Gomes, {\it {An introduction to higher-form symmetries}},  {\em SciPost Phys. Lect. Notes} {\bf 74} (2023) 1, [\href{http://arxiv.org/abs/2303.01817}{{\tt arXiv:2303.01817}}].

\bibitem{Bhardwaj:2023kri}
L.~Bhardwaj, L.~E. Bottini, L.~Fraser-Taliente, L.~Gladden, D.~S.~W. Gould, A.~Platschorre, and H.~Tillim, {\it {Lectures on generalized symmetries}},  {\em Phys. Rept.} {\bf 1051} (2024) 1--87, [\href{http://arxiv.org/abs/2307.07547}{{\tt arXiv:2307.07547}}].

\bibitem{Schafer-Nameki:2023jdn}
S.~Schafer-Nameki, {\it {ICTP lectures on (non-)invertible generalized symmetries}},  {\em Phys. Rept.} {\bf 1063} (2024) 1--55, [\href{http://arxiv.org/abs/2305.18296}{{\tt arXiv:2305.18296}}].

\bibitem{Shao:2023gho}
S.-H. Shao, {\it {What's Done Cannot Be Undone: TASI Lectures on Non-Invertible Symmetries}},  \href{http://arxiv.org/abs/2308.00747}{{\tt arXiv:2308.00747}}.

\bibitem{Brennan:2023mmt}
T.~D. Brennan and S.~Hong, {\it {Introduction to Generalized Global Symmetries in QFT and Particle Physics}},  \href{http://arxiv.org/abs/2306.00912}{{\tt arXiv:2306.00912}}.

\bibitem{Adler:1969gk}
S.~L. Adler, {\it {Axial vector vertex in spinor electrodynamics}},  {\em Phys. Rev.} {\bf 177} (1969) 2426--2438.

\bibitem{Bell:1969ts}
J.~S. Bell and R.~Jackiw, {\it {A PCAC puzzle: $\pi^0 \to \gamma \gamma$ in the $\sigma$ model}},  {\em Nuovo Cim. A} {\bf 60} (1969) 47--61.

\bibitem{Choi:2022jqy}
Y.~Choi, H.~T. Lam, and S.-H. Shao, {\it {Noninvertible Global Symmetries in the Standard Model}},  {\em Phys. Rev. Lett.} {\bf 129} (2022), no.~16 161601, [\href{http://arxiv.org/abs/2205.05086}{{\tt arXiv:2205.05086}}].

\bibitem{Cordova:2022ieu}
C.~Cordova and K.~Ohmori, {\it {Noninvertible Chiral Symmetry and Exponential Hierarchies}},  {\em Phys. Rev. X} {\bf 13} (2023), no.~1 011034, [\href{http://arxiv.org/abs/2205.06243}{{\tt arXiv:2205.06243}}].

\bibitem{Damia:2022bcd}
J.~A. Damia, R.~Argurio, and E.~Garcia-Valdecasas, {\it {Non-invertible defects in 5d, boundaries and holography}},  {\em SciPost Phys.} {\bf 14} (2023), no.~4 067, [\href{http://arxiv.org/abs/2207.02831}{{\tt arXiv:2207.02831}}].

\bibitem{Karasik:2022kkq}
A.~Karasik, {\it {On anomalies and gauging of U(1) non-invertible symmetries in 4d QED}},  {\em SciPost Phys.} {\bf 15} (2023), no.~1 002, [\href{http://arxiv.org/abs/2211.05802}{{\tt arXiv:2211.05802}}].

\bibitem{GarciaEtxebarria:2022jky}
I.~Garc{\'\i}a~Etxebarria and N.~Iqbal, {\it {A Goldstone theorem for continuous non-invertible symmetries}},  {\em JHEP} {\bf 09} (2023) 145, [\href{http://arxiv.org/abs/2211.09570}{{\tt arXiv:2211.09570}}].

\bibitem{Yokokura:2022alv}
R.~Yokokura, {\it {Non-invertible symmetries in axion electrodynamics}},  \href{http://arxiv.org/abs/2212.05001}{{\tt arXiv:2212.05001}}.

\bibitem{Choi:2022fgx}
Y.~Choi, H.~T. Lam, and S.-H. Shao, {\it {Non-invertible Gauss law and axions}},  {\em JHEP} {\bf 09} (2023) 067, [\href{http://arxiv.org/abs/2212.04499}{{\tt arXiv:2212.04499}}].

\bibitem{vanBeest:2023dbu}
M.~van Beest, P.~Boyle~Smith, D.~Delmastro, Z.~Komargodski, and D.~Tong, {\it {Monopoles, scattering, and generalized symmetries}},  {\em JHEP} {\bf 03} (2025) 014, [\href{http://arxiv.org/abs/2306.07318}{{\tt arXiv:2306.07318}}].

\bibitem{Benedetti:2023owa}
V.~Benedetti, H.~Casini, and J.~M. Magan, {\it {ABJ anomaly as a U(1) symmetry and Noether{\textquoteright}s theorem}},  {\em SciPost Phys.} {\bf 18} (2025), no.~2 041, [\href{http://arxiv.org/abs/2309.03264}{{\tt arXiv:2309.03264}}].

\bibitem{Arbalestrier:2024oqg}
A.~Arbalestrier, R.~Argurio, and L.~Tizzano, {\it {Noninvertible axial symmetry in QED comes full circle}},  {\em Phys. Rev. D} {\bf 110} (2024), no.~10 105012, [\href{http://arxiv.org/abs/2405.06596}{{\tt arXiv:2405.06596}}].

\bibitem{Chen:2025buv}
S.~Chen, A.~Cherman, and M.~Neuzil, {\it {Symmetry theta angles and topological Witten effects}},  \href{http://arxiv.org/abs/2507.00220}{{\tt arXiv:2507.00220}}.

\bibitem{Gagliano:2025oqv}
F.~Gagliano, {\it {Quantifying non-invertible chiral symmetry breaking}},  \href{http://arxiv.org/abs/2508.09254}{{\tt arXiv:2508.09254}}.

\bibitem{Cordova:2022fhg}
C.~Cordova, S.~Hong, S.~Koren, and K.~Ohmori, {\it {Neutrino Masses from Generalized Symmetry Breaking}},  \href{http://arxiv.org/abs/2211.07639}{{\tt arXiv:2211.07639}}.

\bibitem{Cordova:2023her}
C.~Cordova, S.~Hong, and L.-T. Wang, {\it {Axion domain walls, small instantons, and non-invertible symmetry breaking}},  {\em JHEP} {\bf 05} (2024) 325, [\href{http://arxiv.org/abs/2309.05636}{{\tt arXiv:2309.05636}}].

\bibitem{Cordova:2024ypu}
C.~Cordova, S.~Hong, and S.~Koren, {\it {Noninvertible Peccei-Quinn Symmetry and the Massless Quark Solution to the Strong CP Problem}},  {\em Phys. Rev. X} {\bf 15} (2025), no.~3 031011, [\href{http://arxiv.org/abs/2402.12453}{{\tt arXiv:2402.12453}}].

\bibitem{Delgado:2024pcv}
A.~Delgado and S.~Koren, {\it {Non-invertible Peccei-Quinn symmetry, natural 2HDM alignment, and the visible axion}},  {\em JHEP} {\bf 02} (2025) 178, [\href{http://arxiv.org/abs/2412.05362}{{\tt arXiv:2412.05362}}].

\bibitem{Cardy:1981qy}
J.~L. Cardy and E.~Rabinovici, {\it {Phase structure of Zp models in the presence of a {\ensuremath{\theta}} parameter}},  {\em Nucl. Phys. B} {\bf 205} (1982) 1--16.

\bibitem{Cardy:1981fd}
J.~L. Cardy, {\it {Duality and the {\ensuremath{\theta}} parameter in Abelian lattice models}},  {\em Nucl. Phys. B} {\bf 205} (1982) 17--26.

\bibitem{Shapere:1988zv}
A.~D. Shapere and F.~Wilczek, {\it {Selfdual Models with Theta Terms}},  {\em Nucl. Phys. B} {\bf 320} (1989) 669--695.

\bibitem{Deser:1976iy}
S.~Deser and C.~Teitelboim, {\it {Duality Transformations of Abelian and Nonabelian Gauge Fields}},  {\em Phys. Rev. D} {\bf 13} (1976) 1592--1597.

\bibitem{Witten:1995gf}
E.~Witten, {\it {On $S$-duality in Abelian gauge theory}},  {\em Selecta Math.} {\bf 1} (1995) 383, [\href{http://arxiv.org/abs/hep-th/9505186}{{\tt hep-th/9505186}}].

\bibitem{Colwell:2015wna}
K.~Colwell and J.~Terning, {\it {S-Duality and Helicity Amplitudes}},  {\em JHEP} {\bf 03} (2016) 068, [\href{http://arxiv.org/abs/1510.07627}{{\tt arXiv:1510.07627}}].

\bibitem{Csaki:2010rv}
C.~Csaki, Y.~Shirman, and J.~Terning, {\it {Anomaly Constraints on Monopoles and Dyons}},  {\em Phys. Rev. D} {\bf 81} (2010) 125028, [\href{http://arxiv.org/abs/1003.0448}{{\tt arXiv:1003.0448}}].

\bibitem{Witten:1979ey}
E.~Witten, {\it {Dyons of Charge e theta/2 pi}},  {\em Phys. Lett. B} {\bf 86} (1979) 283--287.

\bibitem{Csaki:2020inw}
C.~Csaki, S.~Hong, Y.~Shirman, O.~Telem, J.~Terning, and M.~Waterbury, {\it {Scattering amplitudes for monopoles: pairwise little group and pairwise helicity}},  {\em JHEP} {\bf 08} (2021) 029, [\href{http://arxiv.org/abs/2009.14213}{{\tt arXiv:2009.14213}}].

\bibitem{Csaki:2020yei}
C.~Cs{\'a}ki, S.~Hong, Y.~Shirman, O.~Telem, and J.~Terning, {\it {Completing Multiparticle Representations of the Poincar{\'e} Group}},  {\em Phys. Rev. Lett.} {\bf 127} (2021), no.~4 041601, [\href{http://arxiv.org/abs/2010.13794}{{\tt arXiv:2010.13794}}].

\bibitem{Hsin:2018vcg}
P.-S. Hsin, H.~T. Lam, and N.~Seiberg, {\it {Comments on One-Form Global Symmetries and Their Gauging in 3d and 4d}},  {\em SciPost Phys.} {\bf 6} (2019), no.~3 039, [\href{http://arxiv.org/abs/1812.04716}{{\tt arXiv:1812.04716}}].

\bibitem{Laperashvili:1999pu}
L.~V. Laperashvili and H.~B. Nielsen, {\it {Dirac relation and renormalization group equations for electric and magnetic fine structure constants}},  {\em Mod. Phys. Lett. A} {\bf 14} (1999) 2797, [\href{http://arxiv.org/abs/hep-th/9910101}{{\tt hep-th/9910101}}].

\bibitem{Csaki:2024ajo}
C.~Cs{\'a}ki, R.~Ovadia, O.~Telem, J.~Terning, and S.~Yankielowicz, {\it {Abelian instantons and monopole scattering}},  {\em JHEP} {\bf 11} (2024) 165, [\href{http://arxiv.org/abs/2406.13738}{{\tt arXiv:2406.13738}}].

\bibitem{GarciaGarcia:2025uub}
I.~Garcia~Garcia, M.~Kongsore, and K.~Van~Tilburg, {\it {Dyon Loops and Abelian Instantons}},  \href{http://arxiv.org/abs/2506.14867}{{\tt arXiv:2506.14867}}.

\bibitem{Fan:2021ntg}
J.~Fan, K.~Fraser, M.~Reece, and J.~Stout, {\it {Axion Mass from Magnetic Monopole Loops}},  {\em Phys. Rev. Lett.} {\bf 127} (2021), no.~13 131602, [\href{http://arxiv.org/abs/2105.09950}{{\tt arXiv:2105.09950}}].

\bibitem{Fujikawa:1979ay}
K.~Fujikawa, {\it {Path Integral Measure for Gauge Invariant Fermion Theories}},  {\em Phys. Rev. Lett.} {\bf 42} (1979) 1195--1198.

\bibitem{Fujikawa:1980eg}
K.~Fujikawa, {\it {Path Integral for Gauge Theories with Fermions}},  {\em Phys. Rev. D} {\bf 21} (1980) 2848. [Erratum: Phys.Rev.D 22, 1499 (1980)].

\bibitem{Kim:1979if}
J.~E. Kim, {\it {Weak Interaction Singlet and Strong CP Invariance}},  {\em Phys. Rev. Lett.} {\bf 43} (1979) 103.

\bibitem{Shifman:1979if}
M.~A. Shifman, A.~I. Vainshtein, and V.~I. Zakharov, {\it {Can Confinement Ensure Natural CP Invariance of Strong Interactions?}},  {\em Nucl. Phys. B} {\bf 166} (1980) 493--506.

\bibitem{Sokolov:2022fvs}
A.~V. Sokolov and A.~Ringwald, {\it {Electromagnetic Couplings of Axions}},  \href{http://arxiv.org/abs/2205.02605}{{\tt arXiv:2205.02605}}.

\bibitem{Sokolov:2023pos}
A.~V. Sokolov and A.~Ringwald, {\it {Generic Axion Maxwell Equations: Path Integral Approach}},  {\em Annalen Phys.} {\bf 536} (2023), no.~1 2300112, [\href{http://arxiv.org/abs/2303.10170}{{\tt arXiv:2303.10170}}].

\bibitem{Heidenreich:2023pbi}
B.~Heidenreich, J.~McNamara, and M.~Reece, {\it {Non-standard axion electrodynamics and the dual Witten effect}},  {\em JHEP} {\bf 01} (2024) 120, [\href{http://arxiv.org/abs/2309.07951}{{\tt arXiv:2309.07951}}].

\bibitem{Csaki:2024plt}
C.~Cs{\'a}ki, R.~Ovadia, M.~Ruhdorfer, O.~Telem, and J.~Terning, {\it {The Seiberg-Witten Axion}},  \href{http://arxiv.org/abs/2411.15312}{{\tt arXiv:2411.15312}}.

\bibitem{Csaki:work}
C.~Cs{\'a}ki, R.~Ovadia, M.~Ruhdorfer, O.~Telem, and J.~Terning, {\it {work in progress}}, .

\bibitem{Bott:1982xhp}
R.~Bott and L.~W. Tu, {\em {Differential Forms in Algebraic Topology}}.
\newblock Springer, 1982.

\bibitem{Nakahara:2003nw}
M.~Nakahara, {\em {Geometry, topology and physics}}.
\newblock 2003.

\bibitem{Terning:2018lsv}
J.~Terning and C.~B. Verhaaren, {\it {Dark Monopoles and $SL(2,\mathbb Z)$ Duality}},  {\em JHEP} {\bf 12} (2018) 123, [\href{http://arxiv.org/abs/1808.09459}{{\tt arXiv:1808.09459}}].

\end{thebibliography}\endgroup

%%%%%%%%%%%%%%%%%%%%%%%%%%%%%%%%%%%%%%%%%%%%%%%%%%%%%%%%%%%%%%%%%%%%%%%%%%%%%%%%%%%%%%%%%%%%%%%%%%%%

\end{document}